\documentclass[11pt,a4wide]{article}

\usepackage{fullpage}

\usepackage{graphics,graphicx}
\usepackage[dvips]{epsfig}
\usepackage{amsmath,amssymb,amsfonts}

\newtheorem{theorem}{Theorem}[section]
\newtheorem{lemma}{Lemma}[section]
\newtheorem{corollary}{Corollary}[section]
\newtheorem{claim}{Claim}[section]

\newtheorem{definition}{Definition}[section]

\newcommand{\qed}{\hfill $\Box$ \bigbreak}
\newenvironment{proof}{\noindent {\bf Proof.}}{\qed}

\newcommand{\cP}{{\cal P}}
\newcommand{\cC}{{\cal C}}
\newcommand{\cV}{{\cal V}}

\newcommand{\mad}{\mbox{\rm MAD}}
\newcommand{\mav}{\mbox{\rm MAV}}
\newcommand{\id}{\mbox{\rm Id}}
\newcommand{\inp}{\mbox{\bf w}}
\newcommand{\certif}{\mbox{\bf x}}
\newcommand{\teamsize}{\mbox{\tt teamsize}}
\newcommand{\nbnode}{\mbox{\tt \#nodes}}
\newcommand{\tree}{\mbox{\tt tree}}
\newcommand{\path}{\mbox{\tt path}}
\newcommand{\leader}{\mbox{\tt leader}}
\newcommand{\odd}{\mbox{\tt odd}}
\newcommand{\map}{\mbox{\tt map}}
\newcommand{\treesize}{\mbox{\tt treesize}}

\newcommand{\leaf}{\mbox{\tt leaf}}
\newcommand{\Omegastar}{\Omega}
\newcommand{\quotient}{\mbox{\tt quotient}}
\newcommand{\degree}{\mbox{\tt degree}}
\newcommand{\cycle}{\mbox{\tt cycle}}
\newcommand{\sun}{\mbox{\tt sun}}

\renewcommand{\gather}{\mbox{\sc gather}}
\newcommand{\rdv}{\mbox{\sc rdv}}
\newcommand{\token}{\mbox{\sc token}}
\newcommand{\omeg}{\mbox{\sc omega}}

\begin{document}
%%%%%%%%%%%%%%%%%%%%%%%%%%%%%%%%%%%%%%%%%%%%%%%%%%%%

\parskip   1ex

\title{{\bf Decidability Classes for Mobile Agents Computing}}

\author{
Pierre Fraigniaud\thanks{CNRS and Universit\'e Paris Diderot, France.
E-mail: Pierre.Fraigniaud@liafa.jussieu.fr.
Part of this work was done during
this author's visit at the Research Chair in Distributed Computing of
the Universit\'{e} du Qu\'{e}bec en Outaouais. Additional supports from ANR projects ALADDIN and PROSE, and INRIA project GANG.}
\and
Andrzej Pelc\thanks{D\'{e}partement d'informatique, Universit\'{e} du Qu\'{e}bec en Outaouais,
Gatineau, Qu\'{e}bec J8X 3X7,
Canada. E-mail: pelc@uqo.ca.
Part of this work was done during
this author's visit at LIAFA, Universit\'e Paris Diderot, France. Supported in part by NSERC discovery grant 
and by the Research Chair in Distributed Computing of
the Universit\'{e} du Qu\'{e}bec en Outaouais.}
}

\date{ }
\maketitle

\begin{abstract}

We establish a classification of decision problems that are to be solved by mobile agents operating in unlabeled graphs, using a deterministic protocol.
The classification is with respect to the ability of a team of agents to solve the problem, possibly with the aid of additional information.
In particular, our focus is on studying differences between the {\em decidability} of a decision problem by agents and its {\em verifiability}
when a {\em certificate} for a positive answer is provided to the agents. 
We show that the class $ \mav$ of {\em mobile agents verifiable} problems is much wider than the class
$\mad$ of {\em mobile agents decidable} problems. 
Our main result shows that there exist natural $\mav$-complete problems: the most difficult problems in this class, to which all problems in $\mav$
are reducible. Our construction of a $\mav$-complete problem involves two main ingredients in mobile agents computability: the topology of 
the quotient graph and the number of operating agents. Beyond the class $\mav$ 
we show that, for a single agent, three natural oracles yield a strictly increasing chain of relative decidability classes.
\vspace{2ex}

\end{abstract}

\vfill

\vfill

\thispagestyle{empty}
\setcounter{page}{0}
\pagebreak

%%%%%%%%%%%%%%%%%%%%%%%%%%%%%%%%%%%%%%%%%%%%%%%%%%%%%%%%%%%
\section{Introduction}
%%%%%%%%%%%%%%%%%%%%%%%%%%%%%%%%%%%%%%%%%%%%%%%%%%%%%%%%%%%

%--------------------------------------------------------------
\subsection{The context and the problem}
%--------------------------------------------------------------

Algorithmic aspects of mobile agents computing have received growing attention in the recent years. Two scenarios are usually 
studied in this context: mobile entities operate either in the plane, in which case they model, e.g., physical robots executing such tasks as
gathering (rendezvous) \cite{anderson98a,CFPS,fpsw} or pattern formation \cite{DFSY}, or they move 
in a connected graph that models a communication network. In the latter case the mobile entities represent, e.g., software agents.
Computation tasks assigned to mobile agents operating in graphs range from graph exploration \cite{AH,DP}, used, e.g., in network maintenance, detecting faults, or 
searching for information in distributed databases, to gathering in one node \cite{alpern99,CCGL,TSZ07}, in order to exchange data acquired by agents or
to coordinate further actions. Algorithmic problems in mobile agents computing concern both the feasibility of a given task and
its efficiency in terms of the time of accomplishing the task or the memory needed by agents to complete it. Feasibility of gathering in the plane was studied, e.g, in
\cite{CFPS,fpsw} and in graphs in \cite{CLP}. Exploration time for mobile agents in graphs was studied, e.g., in  \cite{AH,DP} in the case of one agent and in
\cite{BS,FGKP} for several agents.  The time of deterministic rendezvous was the subject, e.g.,  of \cite{CLP,DFKP,KM,TSZ07}, while that of randomized rendezvous
was investigated, e.g, in \cite{alpern02a,alpern99}. See also the book \cite{alpern02b} partly devoted to the efficiency of randomized rendezvous.
Memory needed for graph exploration by a single agent was investigated, e.g., in \cite{FI04,GPRZ,R08} and memory needed for rendezvous was the object of study
in \cite{CKP,FP} for the deterministic scenario and in \cite{KKM} for the randomized scenario.

In this paper we present a different perspective on deterministic mobile agents computing in graphs. We are interested in decision problems that may be stated in
this computing environment. These problems may concern various properties of the initial configuration of agents in the graph, e.g., 
``Is the graph a tree?'', ``Are there more than three agents in the graph?'', ``Are the agents located at distance at least $k$?'' The decision has to be made collectively by the agents
and satisfy the following condition: If the answer is ``yes'', then all agents must answer ``yes'', and if the answer is ``no'', then at least one agent must answer ``no''.
We require that all agents eventually decide. (In fact, all our deciding algorithms satisfy a stronger condition: all agents are unanimous also in the negative case.) 
Note that a major difficulty in making a decision by a team of mobile agents is that none of the agents is {\em a priori} provided with the initial configuration on whose properties they have to decide.

Our aim is to classify decision problems with respect to the ability of a team of agents to solve the problem,  
possibly with the aid of additional information.
We focus on studying differences between the {\em decidability} of a decision problem by agents and its {\em verifiability}
when a {\em certificate} for a positive answer is provided to the agents. 

%--------------------------------------------------------------
\subsection{Our results}
%--------------------------------------------------------------

We show that the class $ \mav$ of {\em mobile agents verifiable} problems is much wider than the class $\mad$ of {\em mobile agents decidable} problems. In particular, we show that it contains an infinite antichain with respect to the reducibility relation.

Our main result shows that $\mav$ contains a natural complete problem:
a problem to which all problems in $\mav$
are reducible. The problem is composed of two ``orthogonal'' parts
that are in the core of mobile agents computing: one concerns the number of operating agents and the other concerns the quotient
graph\footnote{Informally,  the quotient is taken with respect to the equivalence relation between 
nodes that have the same ``view'' of the graph. See Section~\ref{sec:quotient} for a precise definition of the quotient graph.}  of the initial configuration. More precisely, our $\mav$-complete problem involves the problems $\teamsize$ and $\quotient$,
where $\teamsize$ is the problem to decide whether the number of agents is larger than a given positive integer $k$, and
$\quotient$ is the problem to decide whether the quotient of the graph where the agents operate
is different from a given graph $H$.

We also look beyond the class of mobile agents verifiable problems,  showing that,  for a single agent, three 
natural oracles (decision problems whose solution is given as a black box) 
yield a strictly increasing chain of relative decidability classes. 

%%%%%%%%%%%%%%%%%%%%%%%%%%%%%%%%%%%%%%%%%%%%%%%%%%%%%%%%%%%
\section{Mobile agents computing and complexity classes}
%%%%%%%%%%%%%%%%%%%%%%%%%%%%%%%%%%%%%%%%%%%%%%%%%%%%%%%%%%%

%--------------------------------------------------------------
\subsection{Mobile agents model} 
%--------------------------------------------------------------

Agents operate in simple undirected connected graphs without node labels.  
Agents cannot leave any marks at visited nodes. The first assumption
is motivated by the fact that nodes may refuse to reveal their identities, e.g., for security reasons, or limited sensory capabilities of the agents
may prevent them from perceiving these identities. The reason for  the second assumption is that nodes may have no facilities 
(whiteboards) allowing to leave marks, or such marks may
be destroyed between visits of the agents and thus are unreliable. 
By contrast, in order to allow the agents to move in the network,  we have to assume that ports at every node are distinguishable for the agents. 
If an agent were unable to locally distinguish ports at a node, it may have even been unable to
visit all neighbors of a node of degree at least 3. Indeed, after visiting the second neighbor, the agent cannot distinguish
the port leading to the first visited neighbor from the port leading to the unvisited one.  Thus an adversary may always force an agent to avoid all
but two edges incident to such a node, thus effectively precluding exploration. Hence we assume that
a node of degree $d$ has ports $1,\dots, d$ corresponding to the incident edges. Ports at each node can be perceived by an agent visiting this node, but there is no coherence assumed between port labelings at different nodes. From now on, a graph will always mean a connected graph without node labels but with port labels.
For a graph $G$, we denote by $V(G)$ the set of nodes of $G$, and call $|V(G)|$ the {\em size} of $G$. Feasibility of various computing tasks in unlabeled graphs is a classic object of study: see, e.g., \cite{A,BV,YK3}.
We define below the inputs of the computing tasks that are considered in this paper.

An {\em initial configuration} is a quadruple  $(G,S,\id,\inp)$, where $G$ is a graph, $S \subseteq V(G)$ is a non-empty multiset, $\id$ is a one-to-one function from $S$ to the set of positive integers, and $\inp$ is a function from $S$ to the set $\{0,1\}^*$ of binary strings.

The set $S$ is interpreted as the set of nodes hosting agents at the start. It is actually a multiset since there might be initially more than one agent at a node of $S$, and thus there can be more than one occurrence of a same node in $S$. For $s \in S$, the value $\id(s)$ is the identity of the agent hosted by node $s$, and the value $\inp(s)$ is the input of this agent. (If there is more than one occurrence of $s$, each of them receives a different identity, and a non-necessarilly different input). Initially, an  agent does not have any a priori knowledge of the initial configuration apart from its own identity and its own input. For many problems, the inputs of the agents are all identical, in which case the input $\inp$ is simply denoted by a binary string $w\in\{0,1\}^*$. (We denote by $\epsilon$ the empty binary string). Note that although $\id$ and $\inp$ are defined as ``functions'', there is no computability issue involved in the definition because one does not ask agents to compute their identities and inputs. In fact, $\id$ and $\inp$ can also be viewed as vectors of $|S|$ coordinates. 
 
Agents are abstract state machines with distinct identities and unlimited memory. 
Agents start simultaneously and move in synchronous rounds:
in each round an agent can stay in the current node or move to an adjacent node. 
When coming to a node, an agent recognizes the entry port number,  the degree of the node and the identities of all agents currently located at this node (if any). When two agents meet at the same node in the same round, they can exchange
all information they currently have. 
On this basis, together with the content of its memory, the agent computes the port number by which it leaves the node in the next round, or decides to stay in the current node. Note that the assumption about synchrony is made only to simplify presentation: all our study can be carried out for asynchronous
agents that are allowed to meet not only at a node but also inside an edge. In this case, however, there are additional technicalities needed to model an adversary
 representing asynchrony (cf. \cite{CLP}). Also meeting inside an edge is not a natural assumption from the point of view of applications for software agents, hence we restrict attention
to the synchronous scenario.

%--------------------------------------------------------------
\subsection{Decision and verification problems}
%--------------------------------------------------------------

We define decision and verification problems, as well as their relative corresponding classes, in the context of computing with mobile agents. 

\begin{definition}\label{def:decisinpb}
A {\em decision problem} is a set $\Pi$ of initial configurations such that
\begin{enumerate}
\item there exists an algorithm which, given any initial configuration $(G,S,\id,\inp)$, decides whether \\$(G,S,\id,\inp)\in \Pi$;
\item $\Pi$ is closed under automorphisms in the following sense: if $\alpha$ is an automorphism of a graph $G$ preserving port numbers, then: $$(G,S,\id,\inp) \in \Pi \iff (G,\alpha(S),\id \circ \alpha ^{-1},\inp\circ \alpha ^{-1}) \in \Pi.$$
\end{enumerate}
We denote by $\Delta$ the class of all decision problems for mobile agents. 
\end{definition}

Condition~1 expresses the fact that we are interested in decidable problems only, so as to identify decidable problems that cannot be decided in the framework of mobile agents computing. (Undecidable problems obviously remain undecidable in this framework).
Condition~2 expresses the fact that, since nodes
of the graph are not labeled, no distinction can be made between two configurations that can be carried on each other by a graph  automorphism preserving
port numbers and agents' identities and inputs. 

A typical example of decision problems is $\teamsize=\{(G,S,\id,k):|S|>k\}$, where agents are provided with the integer $k$, and must decide whether there are more than $k$ agents in the graph. Another example is $\nbnode=\{(G,S,\id,n):|V(G)|=n\}$, where agents are provided with the integer $n$, and must decide whether the size of the graph is $n$. A third example is the (input-free) decision problem $\tree=\{(G,S,\id,\epsilon): G \; \mbox{is a tree}\}$. This latter problem has a natural variant $\treesize=\{(G,S,\id,n): G \; \mbox{is a tree and $|V(G)|=n$}\}$. Finally, the problem $\leader = \{(G,S,\id,\inp): ||\inp||_1=1\}$, for the agents, each receiving a single bit as input, consists in deciding whether there is a unique agent with input~1, with all the others having input~0. 

A decision problem $\Pi$ is {\em mobile agents decidable}, if there exists a protocol for agents such that any team  of agents provided with arbitrary distinct identities $\id$, arbitrary input $\inp$, and executing this protocol in any graph $G$ starting from positions $S$ satisfies that all agents eventually make a decision ``yes'' or ``no'' according to the following {\em decision} property:

\begin{itemize}
\item if $(G,S,\id,\inp) \in \Pi$, then all agents decide ``yes'';

\item if $(G,S,\id,\inp) \not \in \Pi$, then at least one agent decides ``no''. 
\end{itemize}

Note that there is a significant difference between usual decidability of a decision problem (expressed by Condition~1 of Definition~\ref{def:decisinpb})
and mobile agents decidability: no agent knows {\em a priori} the initial configuration (but only its own identity and its input), hence it has either to learn it during the execution of the protocol or make its decision without full knowledge of the initial configuration. 
The class of all mobile agents decidable problems is denoted by $\mad$. 

It may appear at a first glance that  $\mad$ is the class of problems that involve some form of ``locality'', since problems like deciding whether there exists a node $u$ satisfying some property ${\cal{P}}_u$, or deciding whether all nodes satisfy some property ${\cal{P}}$ are usually not in $\mad$ because it is hard for an agent to visit all nodes of an arbitrary graph. Nevertheless, the ``global'' problem $\odd=\{(G,S,\id,\epsilon): \mbox{the initial node has odd degree, and there is another node  of odd degree}\}$ is obviously in $\mad$. A non trivial problem in $\mad$ is $\treesize$ with a protocol for each agent that consists in performing a DFS for $2(n-1)$ steps, and to draw a map of the visited graph: if the agent is back at its original position on the map with no edge unvisited on the map, then it decides ``yes'', otherwise it decides ``no''. On the other hand, neither $\nbnode$ nor $\tree$ belongs to $\mad$ (even $\path=\{(G,S,\id,\epsilon): G\; \mbox{is a path}\}$ is not in $\mad$ because, informally, a single agent cannot distinguish a long path from a cycle\footnote{Indeed, place one agent in an infinite path, whose edges are consistently labeled~1 rightward, and~2 leftward. Since the agent has to decide, let $t$ be the number of rounds performed by the agent before it decides. The agent cannot distinguish the case in which it is placed at the central node of a (consistently labeled) path with $2t+3$ nodes from the case in which it is placed in a  (consistently labeled)  cycle.}). 

For a decision problem, deciding it is different from {\em verifying} it. In the latter case agents are presented with a {\em certificate}, if the answer to the problem is ``yes'',
and using a verifying protocol have to decide ``yes''. Moreover, they have to be immune to ``cheating'' them: no certificate can induce them to decide ``yes'', if the real answer is ``no''.  This framework is reminiscent of the distinction between classes $\mbox{NP}$ and $\mbox{P}$ in the theory of complexity \cite{CLR}. 

The notion of verification is formally defined as follows.
A {\em certificate} $\certif$ is a function from the multiset $S$ of initial positions to $\{0,1\}^*$. That is, agent $s$ receives the ``partial'' certificate $\certif(s)$.  A  {\em verification protocol} is executed by agents provided with certificate $\certif$. (In the sequel, most of our protocols will use the same certificate $x\in\{0,1\}^*$ for all agents). A decision problem $\Pi$ is {\em mobile agents verifiable}, if there exists a verification protocol for agents such that any team of agents provided with arbitrary distinct identities $\id$, arbitrary inputs $\inp$, and executing this protocol in any graph $G$ starting from positions $S$ satisfies that all agents eventually make a decision ``yes'' or ``no'' according to the following {\em verification} property:

\begin{itemize}
\item if $(G,S,\id,\inp) \in \Pi$, then there exists a certificate $\certif$ such that all agents decide ``yes'';

\item if $(G,S,\id,\inp) \not \in \Pi$, then, for every certificate $\certif$, at least one agent decides ``no''.
\end{itemize}

The class of all mobile agents verifiable problems is denoted by $\mav$. 

By definition, $\mad\subseteq\mav$. An example of a problem that is not mobile agents decidable but is mobile agents verifiable is $\path$. We have seen that $\path\notin\mad$. However, $\path\in\mav$ by using a certificate interpreted as the number of nodes in the path. Similarly, $\tree\in\mav$ by using the same certificate, and performing a verifying protocol similar to the one used for proving $\treesize\in\mad$. The problem $\leaf=\{(G,S,\id,\epsilon): \exists u\in V(G), \deg(u)=1\}$ is in $\mav$: for each $s\in S$, provide agent $s$ with certificate $\certif(s)$ describing a path from $s$ to a leaf (i.e., the list of port numbers along a path from $s$ to a leaf). 

We define the relation of reducibility between decision problems in the usual way. For the sake of simplicity, we restrict our attention to reductions to a subclass of decision problems. This will be proved to be sufficient for the purpose of this paper. We say that a decision problem is \emph{uniform} if and only if all its instances are initial configurations $(G,S,\id,\inp)$ such that $\inp(u)=\inp(v)$ for every two nodes $u,v\in S$. For instance, $\nbnode$ is uniform. 

A problem $\Pi$ is {\em reducible} to a uniform problem $\Pi'$, denoted $\Pi \preceq \Pi'$, if there exists a protocol for mobile agents to decide $\Pi$, using a black box procedure deciding $\Pi'$ that can be used an arbitrary finite number of times. 

In the above definition, using the black box deciding $\Pi'$ means feeding the black box with arbitrary input binary strings. In other words, assuming that $(G,S,\id,\inp)$ is the initial configuration, an agent uses the black box deciding $\Pi'$ by calling the black box with its input $w\in\{0,1\}^*$, and the black box decides whether  $(G,S,\id,w)\in\Pi'$. For instance, $\tree\preceq\nbnode$.  To see why, consider the protocol consisting in successively calling a black box deciding $\nbnode$ with input integers $1,2,\dots$. The black box eventually answers ``yes" when the tested integer $n$ is the size of the graph $G$ in which the agents are operating. Once the size of the graph $G$ is known, deciding whether $G$ is a tree is achieved using the aforementioned protocol deciding $\treesize$. 

\begin{figure}[tb]
\begin{center}
\includegraphics[width=16cm]{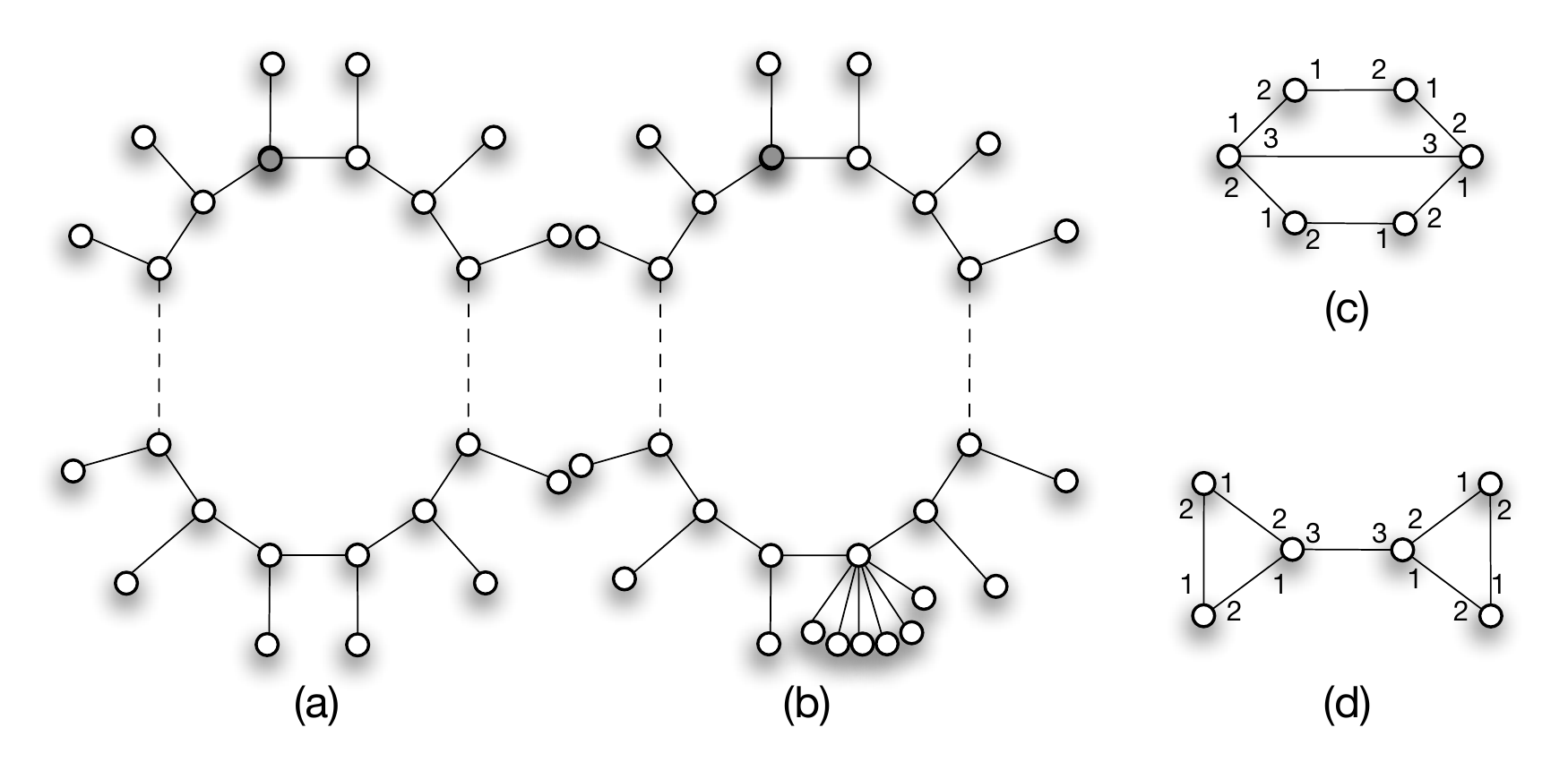}
\caption{Graphs (a) and (b) cannot be distinguished by a single agent (clockwise edges are labeled~1, and counterclockwise edges are labeled~2); Graphs (c) and (d) have the same size and same quotient, but are not isomorphic. }
\label{fig:proof}
\end{center}
\end{figure}

Observe that not only do there exist decision problems that are mobile agents verifiable and not mobile agents decidable, but in fact the class $\mav \setminus \mad$ is quite large. This is shown in the following example. Let us consider the following family of (input-free) decision problems. For every $k\geq 1$, let $\degree_k=\{(G,S,\id,\epsilon): \exists u\in V(G), \deg(u)=k\}.$ Thus, $\degree_k$ is the problem of whether the underlying graph contains a node of degree $k$. All problems $\degree_k$ are in $\mav$, using a certificate which is, for each agent, a path leading the agent from its original position to a node of degree $k$. The problems $\degree_k$ form an infinite antichain with respect to the reducibility relation. Indeed, $\degree_k\not\preceq \degree_{k'}$ for all $k'\neq k$. To see why, consider the graphs of Figures~\ref{fig:proof}~(a) and~(b) where the grey node is the starting position of the unique agent. This agent cannot distinguish these two graphs if their diameters are large enough because the two graphs are identical at bounded distance from the starting position: the agent will not visit the nodes whose degrees differ in the two graphs, prior to making its decision, if these nodes are far enough from the starting positions. 

Our main result is related to the following standard concept: For a class $\cC$ of decision problems, the problem $\Pi$ is $\cC$-{\em complete}, if it belongs to $\cC$ and if any
problem in the class $\cC$ is reducible to $\Pi$.

The main contribution of the paper is to show that there is a natural $\mav$-complete problem, and this is the purpose of the next section. 

%%%%%%%%%%%%%%%%%%%%%%%%%%%%%%%%%%%%%%%%%%%%%%%%%%%%%%%%%%%
\section{The hardness of problems in $\mav$}
\label{sec:hardness}
%%%%%%%%%%%%%%%%%%%%%%%%%%%%%%%%%%%%%%%%%%%%%%%%%%%%%%%%%%%

In this section, we establish our main result, stating that there is a natural $\mav$-complete problem. This problem, denoted by $\Omegastar$,  involves two main components. One is $\teamsize$, the problem consisting in deciding whether the size of the agents' team is larger than a given value, and the other is $\quotient$, the problem consisting in deciding whether the quotient of the graph in which the agents are operating is different from a given graph. We start the section by defining  the problem $\quotient$, which requires defining the notion of the quotient graph.

%--------------------------------------------------------------
\subsection{The problem $\quotient$}
\label{sec:quotient}
%--------------------------------------------------------------

In order to define the problem $\quotient$, we rely on the following important notion introduced in \cite{YK3}. 
Let $G$ be a graph and $v$ a node of $G$. The {\em view} from $v$ is the infinite rooted tree $\cV(v)$ with labeled
ports, defined recursively as follows. The truncated view of $v$ at depth $0$,  $\cV^{(0)}(v)$, is a single node. Given the truncated views at depth $t\geq 0$,  $\cV^{(t)}(u)$, of every node $u$, we define $\cV^{(t+1)}(v)$. $\cV^{(t+1)}(v)$ has its root $x$ corresponding to $v$. For every node $v_i$, $i=1,\dots ,d$, adjacent to $v$ in $G$, there is a child $x_i$ in $\cV^{(t+1)}(v)$ such that the port number at $v$ (resp., at $v_i$) corresponding to edge $\{v,v_i\}$ is the same as the port number at $x$ (resp., at $x_i$) corresponding to edge $\{x,x_i\}$. Node $x_i$, for $i=1,\dots ,d$, 
is then set as the root of the truncated view $\cV^{(t)}(v_i)$ at depth $t$ from $v_i$. The resulting tree is  $\cV^{(t+1)}(v)$. The view $\cV(v)$ from $v$ is the infinite tree
rooted at a node $x$ whose every subtree of depth $t\geq 0$, rooted at $x$, is isomorphic to $\cV^{(t)}(v)$. 
The following result is proved in \cite{No}.

\begin{theorem}\label{trunc}
For every $n$-node graph, and for every two nodes $u$ and $v$ of the graph, if $\cV^{n-1}(u)=\cV^{n-1}(v)$ then $\cV(u)=\cV(v)$. 
\end{theorem}

The  notion of view is used in \cite{YK3} to define one of the crucial concepts in computations in unlabeled graphs. 
Let $G$ be a graph. The {\em quotient graph} of $G$, denoted by $\widehat{G}$, is a (not necessarily simple) graph
defined as follows. Every node of  $\widehat{G}$ corresponds to a maximal set of nodes of $G$ which have the same view. For all  (possibly equal) nodes $x,y\in V(\widehat{G})$  corresponding to two sets $U,V\subseteq V(G)$, respectively, there is an edge between
$x$ and $y$ with labels $p$ at $x$ and $q$ at $y$, if there exists an edge $\{u,v\}$ in $G$ with $u \in U$, $v \in V$ and with ports $p$ at $u$ and $q$ at $v$.  
It follows from Theorem \ref{trunc} that, for every node $v \in V(G)$, computing the truncated view of $G$ from $v$ at depth at least $2(n-1)$ suffices to construct $\widehat{G}$. 
Note that, as opposed to graphs in which agents operate, a quotient graph can have self-loops (whose both ports may have the same number) and multiple edges. It follows from \cite{YK3} that computing truncated views of all nodes of $G$ is the maximum information that can be obtained from exploring $G$ by a single agent. This is summarized by the following lemma.  

\begin{lemma} \label{lem:idexe}
Let $G$ and $H$ be two graphs, and assume that $\widehat{G}=\widehat{H}=Q$. Consider a single agent performing a protocol starting from node $v$ of $G$ and from node $w$ of $H$, where $v$ and $w$ correspond to the same node of $Q$. Assume that the agent is initially aware only of its identity and of its input string. Then the execution of the protocol is identical in $G$ and in $H$.
\end{lemma}

Truncated views being the maximum information that can be obtained from exploring a graph by a single agent, the quotient graph is, intuitively,  the ultimate information that can be gained about the graph from its exploration by a single agent. We are interested in the decision problem  $\quotient=\{(G,S,\id,H): \widehat{G}\neq H\}.$ Thus $\quotient$ is the problem to decide whether the quotient graph of the graph from an initial configuration is \emph{different} from a given graph. 

%--------------------------------------------------------------
\subsection{A $\mav$-complete problem}
%--------------------------------------------------------------

In order to state our main result we define the following {\em product} of decision problems. Let $\Pi_i$, $i=1,\dots,k$, be $k$ decision problems. We define the decision problem $\Pi=\Pi_1\times\Pi_2\times\dots\times\Pi_k$ as follows: 
$$(G,S,\id,(i,\inp))\in\Pi \iff \Big ( 1\leq i \leq k \; \mbox{and} \; (G,S,\id,\inp)\in\Pi_i  \Big )~.$$

We are now ready to state our main result: 

\begin{theorem}\label{theo:main}
The decision problem $\Omegastar= \teamsize \times \quotient$ is $\mav$-complete.
\end{theorem}

\begin{proof}
First, we prove that $\Omegastar\in\mav$. To establish this, we describe a verifying protocol whose certificate $x$ is the same for all agents, and is interpreted as the size $n$ of the network $G$ in which these agents are operating. Let $(G,S,\id,(i,\inp))$ be an initial configuration for $\Omegastar$, $i\in\{1,2\}$. That is, for all $s$, $\inp(s)=k$ if  $i=1$, where $k$ is a positive integer, and $\inp(s)=H$ otherwise, where $H$ is a graph, and
$$(G,S,\id,(i,\inp))\in\Omegastar  \iff  ( ( i=1 \; \mbox{and} \; |S| = k)  \; \mbox{or} \; ( i=2 \; \mbox{and} \; \widehat{G}\neq H  ) )~.$$

Below is a compact description of the verification protocol performed by each agent.  

\smallbreak

\begin{center}
\fbox{
\begin{minipage}{15cm}\small %\footnotesize

\noindent
\underline{Verification protocol:} 

\vspace{1ex}

\noindent
\hspace*{0.5cm} {\bf if} input is $(1,k)$ {\bf then}  \emph{(the agents must decide whether there are $>k$ of them)} \\
\hspace*{1cm} attempt gathering in a graph with $x$ nodes; \\
\hspace*{1cm} {\bf if} more than $k$ agents gather {\bf then} each of these agents decides ``yes''; \\
\hspace*{1cm} {\bf else} each of them decides ``no'';\\
\hspace*{0.5cm} {\bf else}  \emph{(the input is $(2,H)$ and the agents must decide whether $\widehat{G}\neq H$)} \\
\hspace*{1cm} compute the view of $G$ from the starting position, truncated at distance $2\cdot\max\{x,|V(H)|\}$;\\
\hspace*{1cm} construct the quotient graph $Q$ on the basis of this view;\\
\hspace*{1cm} {\bf if}  $Q \neq H$ {\bf then} decide ``yes'' {\bf else}  decide ``no''. 

\end{minipage}
}
\end{center}

\medbreak

We now detail the protocol, and prove its correctness. Let us first consider the case where the input to $\Omegastar$ has the first term~1, i.e., when the problem to be decided is $\teamsize$. In this context, we make use of the following result, which is folklore in the domain of mobile agents computing. 

\begin{claim}\label{claim:rdv} 
Consider two mobile agents with distinct identities placed initially at arbitrary nodes of a graph unknown to the agents, but whose size $n$ is known to the agents. There exists a computable function, $\tau$, depending on both the size $n$ of the graph and the identity $i$ of each agent, and there exists a rendezvous protocol, $\rdv$, which  guarantees that, after $\tau(n,i)$ rounds in an $n$-node graph, the agent with identity $i$ is back at its starting position, and has met the other agent. 
\end{claim}

This claim has been implicitly proved in several papers (see, e.g., \cite{DFKP} which is aiming at minimizing $\tau$, or \cite{CKP} which aims at minimizing the memory of the agent). We provide a short explicit proof for further references in the text. Let $\tau=\tau(n,i)= 2 (i+1) n^n$. The $\rdv$ protocol works as follows. The agent with identity $i$ executes a consecutive series of $i$ DFS traversals of the graph, at distance at most $n$, and then stops at its initial position. Every DFS requires at most $2 n^n$ rounds in a graph of $n$ nodes. Thus the series of DFS traversals requires at most $2 i n^n$ rounds in a graph of $n$ nodes. Let $i$ and $j$ be the identities of the two agents, $i<j$. Agent $j$ will meet agent $i$ at the starting position of the latter during the $(i+1)$st DFS of the former. This completes the proof of the claim. 

The protocol $\rdv$ is modified  to insure gathering of many agents, as follows. Whenever two or more agents meet, they merge and they carry on the execution of the protocol  $\rdv$ trying to meet the other agents. (Merging means that the agent with the largest label among the merged agents becomes a leader, and the one(s) with smaller label(s) follows that leader, performing identical moves). The decision is made by the agent with identity $i$ after having performed $\tau(x,i_{max})$ rounds, where $i_{max}$ is the maximum of all the identities of the agents with which agent~$i$ has already gathered. If agent $i$ ends up in a group of more than $k$ agents, it decides ``yes'', otherwise it decides ``no''. 

We now show that our verification protocol is correct (still in the case where the input to $\Omegastar$ has the first term~1, i.e., when the problem to be decided is $\teamsize$). Assume that the answer to $\Omegastar$ is ``yes'', i.e., there are more than $k$ agents in the graph. In this case, by Claim~\ref{claim:rdv}, for a certificate $x$ equal to the true size $n$ of the graph, all the agents gather, and thus they are able to count themselves precisely. Hence every agent will decide ``yes'', as desired. On the other hand, if the answer to $\Omegastar$ is ``no'', i.e., there are at most $k$ agents in the graph, then, regardless of the certificate $x$, no agent decides ``yes'' because no agent can meet at least $k$ other agents. 

Let us now consider the case where the input to $\Omegastar$  has first term~2, i.e., the problem to be decided is $\quotient$. More specifically, the agents have to decide, with the help of the certificate $x$, whether the quotient $\widehat{G}$ of the graph in which they operate is different from the input graph $H$. For this purpose, each agent computes the truncated view $\cV^d(s)$ at depth $d=2\cdot\max\{x,|V(H)|\}$ from its original position~$s$. Given $\cV^d(s)$, each agent considers all nodes at depth at most $d/2$ in $\cV^d(s)$, and computes their truncated views at depth $d/2$. That is, for every $v\in \cV^d(s)$ at depth at most $d/2$, the agent considers the subtree $T_v$ of height $d/2$ of $\cV^d(s)$ rooted at $v$.  Given all these truncated views $T_v$, each agent merges the nodes $v$ with the same view $T_v$, from which it computes the quotient graph $Q$. The answer to $\quotient$ is given according to whether the resulting graph $Q$ is isomorphic or not to the input graph $H$. The correctness of this verification protocol is based on the following fact which may be of independent interest, as it is an improvement of the result from~\cite{No} (cf. Theorem~\ref{trunc}). 

\begin{claim}\label{claim:view}
Let $\widehat{G}$ be the quotient graph of $G$, and let $\widehat{n}$ be the size of $\widehat{G}$. For every two nodes $u$ and $v$ of the graph, if $\cV^{\widehat{n}-1}(u)=\cV^{\widehat{n}-1}(v)$ then $\cV(u)=\cV(v)$. As a consequence, for every node $v \in V(G)$, computing the truncated view of $G$ from $v$ at depth at least $2(\widehat{n}-1)$ suffices to construct $\widehat{G}$. 
\end{claim}

The intuition to establish the claim is that, in view of Theorem~\ref{trunc}, computing the truncated view of $\widehat{G}$ at depth at least $2(\widehat{n}-1)$  from some node $v\in V(\widehat{G})$ suffices to reconstruct $\widehat{G}$. The claim then follows since, by Lemma~\ref{lem:idexe}, the behavior and the perception of the agent are identical whether it is placed in $G$ or in $\widehat{G}$. Though this intuition is correct, it is not sufficient to prove the claim because, actually, $\widehat{G}$ has multiple edges and self-loops, and, more importantly, different edges incident to the same node of  $\widehat{G}$ may have identical port numbers at that node. The agent is not designed to operate in such graphs. Yet, by the definition of the quotient graph, two edges with identical port numbers incident to a node $u\in V(\widehat{G})$ necessarily lead to the same node $v\in V(\widehat{G})$. Therefore, an agent enhanced with the ability to navigate in a graph with potentially identical port numbers associated to different edges incident to the same node can compute the view of a node  in $\widehat{G}$, truncated at any given depth. More precisely, let us consider an agent whose perception at a node is the number of incident edges with different port numbers, and assume that whenever the agent has computed an out-going port number $i$, the edge through which it leaves the node is selected arbitrarily among those with port number $i$. Then the behavior and the perception of such an agent are identical whether it is placed in $G$ or in $\widehat{G}$. The claim then follows by application of Theorem~\ref{trunc}. 

We use Claim~\ref{claim:view} to prove the correctness of the verification. First, assume that the answer to $\Omegastar$ is ``yes'', i.e., $\widehat{G}\neq H$. In this case,  for a certificate $x$ equal to the true size $n$ of the graph, we have $d\geq 2n$. Thus, by Theorem~\ref{trunc}, each agent has constructed the quotient $Q=\widehat{G}$. Therefore, it can check that $\widehat{G}\neq H$, hence it decides ``yes'', as desired. Now, assume that the answer to $\Omegastar$ is ``no'', i.e.,  the input is a graph $H$ and $\widehat{G}=H$. Regardless of the certificate $x$, each agent computes the view at distance $d\geq 2 |V(H)|=2\widehat{n}$. By Claim~\ref{claim:view}, we get that the graph $Q$ constructed by the algorithm is the graph $\widehat{G}$. Hence every agent detects that  $\widehat{G}=H$, and returns ``no'', as desired. 

This completes the proof that $\Omegastar\in\mav$. It remains to prove that for every problem $\Pi\in\mav$, we have $\Pi \preceq \Omegastar$. Deciding $\Pi$ using $\Omegastar$ as a black box works as follows. Below is a compact description of the protocol performed by each agent, where $\omeg$ is a procedure allowing agents to decide $\Omegastar$, and $(G,S,\id,\inp)$ is the initial configuration.

\smallbreak

\begin{center}
\fbox{
\begin{minipage}{15cm}\small

\noindent
\underline{Reduction protocol:} 

\vspace{1ex}

\noindent
\hspace*{0.5cm} agents compute the size $k$ of the team using $\omeg$; \\
\hspace*{0.5cm}  {\bf if} $k>1$ {\bf then}  \\
\hspace*{1cm} agents gather at the same node, and they exchange their identities $\id$ and inputs $\inp$; \\
\hspace*{1cm} agents compute a map of the graph $G$ with $S$ identified on the map; \\
\hspace*{1cm} agents decide whether $(G,S,\id,\inp)\in\Pi$; \\
\hspace*{0.5cm} {\bf else}  \emph{(there is a unique agent $s$ with $\id(s)=i$ and $\inp(s)=w$)} \\
\hspace*{1cm} agent $s$ computes a map of $\widehat{G}$ with $\widehat{s}$ identified on the map using $\omeg$;\\
\hspace*{1cm} agent $s$ decides whether $(G,\{s\},i,w)\in\Pi$.

\end{minipage}
}
\end{center}

\smallbreak

More precisely, every agent starts by calling $\omeg$ with input $(1,k)$ for successive values $k=1,2,\dots$ until it gets a ``yes'' answer. When this occurs, every agent knows the size $k$ of the team currently in the graph. The behavior of the agent(s) depends on whether $k=1$ or $k>1$. First assume that $k>1$. In this case, the behavior of the agents is based on the following claim. 

\begin{claim}\label{claim:gather} 
Consider a team of $k>1$ agents in a graph $G$ unknown to these agents. The agents are placed at arbitrary nodes of $G$. Assume that each agent is aware of the size $k$ of the team, but is not aware of the size $n$ of the graph $G$. There exists a protocol $\gather$ that accomplishes gathering of all $k$ agents in the same node. 
\end{claim}

To establish the claim, we show how to use the rendezvous protocol $\rdv$ of Claim~\ref{claim:rdv}  to perform gathering of $k$ agents despite the fact that the agents are not aware of the size $n$ of the graph. This is achieved by trying different values of $n$. The main difficulty is that testing different $n$ may desynchronize the agents, whereas Claim~\ref{claim:rdv}  requires the agents to start at the same time. This desynchronization is caused by both the structure of the graph and the different identities. Let $\tau_{max}(n,b)$ be the maximum number of rounds of $\rdv$ when performed on graphs with $n$ nodes by agents with identites on at most $b$ bits. Protocol $\gather$ enumerates the pairs $(n,b)$, and tests each pair as follows. Each agent performs $\rdv$ as in a graph of $n$ nodes, but stops after $\tau_{max}(n,b)$ rounds. During the $\tau_{max}(n,b)$ rounds, if two or more agents meet, they merge in the sense that the one with the largest label becomes a leader, and the one(s) with smaller label(s) follow(s) that leader. Once the $\tau_{max}(n,b)$ rounds are exhausted, and if the $k$ agents have not yet met, the next pair $(n,b)$ is considered. Eventually, the agents will test a pair $(n,b)$ where $n$ is the size of the graph, and $b$ is an upper bound of all agents' label sizes. In this context, protocol $\rdv$ guarantees that all the agents will meet at the same time at the same node. This completes the proof of the claim. 

By Claim~\ref{claim:gather}, the agents are now assumed to have gathered together at the same node. We show how the agents can then compute a map of $G$ with $S$ identified on the map. We use the following result: 

\begin{claim}\label{claim:map}
Consider an agent equipped with a movable token in a graph $G$ unknown to the agent. There exists a protocol $\token$ for the agent that allows the agent to draw an isomorphic copy of~$G$.
\end{claim}

To establish the claim, let $G$ be the graph in which the agent operates, and let $u$ be the starting node of the agent. We describe the protocol $\token$. For $r\geq 0$, let $B_r$ be the ball centered at $u$, of radius $r$ (i.e., the graph induced by all nodes at distance at most $r$ from $u$), and let $S_r$ be the graph induced by all nodes at distance exactly $r$ from $u$. Initially, the agent only knows $B_0$. Assume that the agent has drawn a map of $B_r$. Note that the agent can navigate in $B_r$ using the map. To get the map of $B_{r+1}$, the agent visits every node $v\in S_r$, and traverses every edge incident to $v$ pending out of $B_r$, reaching every node $w$ at distance exactly $r+1$ from $u$.  In order to draw the map of the edges connecting a node in $S_r$ and a node in $S_{r+1}$, as well as the edges  connecting two nodes in $S_{r+1}$, the agent proceeds as follows. It considers every node $w$ at distance exactly $r+1$ from $u$, and treats each one iteratively. For each $w\in S_{r+1}$, the agent places the token at $w$, and then traverses all edges pending out of $B_r$ to discover other edges incident to the same node $w$. That is, whenever the agent reaches node $w'\in S_{r+1}$, it does the following. If it finds the token at $w'$, then $w'=w$ and a new edge incident to $w$ has been discovered. Otherwise, $w'\neq w$, and then the agent traverses all edges incident to $w'$ to identify possible edges between $w'$ and $w$. When all nodes $w\in S_{r+1}$ have been treated, the agent has a map of $B_{r+1}$. The protocol $\token$ is completed when there are no edges pending out from the current ball $B_r$, in which case the map of $B_r$ is the map of $G$. This completes the proof of the claim.

By Claim~\ref{claim:map}, once the $k>1$ agents have gathered, they can collectively construct a map of the graph (the agent with the largest identity acts as the agent in Claim~\ref{claim:map}, while the others play collectively the role of the token). Note that during the execution of all the operations required to gather and to construct a map of the graph, each agent can trace its movements, and thus not only every agent ends up with a map of the graph, but also the agents know the starting position of every agent. Hence, the agents know the initial configuration $(G,S,\id,\inp)$. Since the problem $\Pi$ is, by definition, decidable, the agents can decide whether $(G,S,\id,\inp)\in\Pi$. This completes the case of a team with more than one agent. 

Now assume that there is a single agent in the graph (i.e., $k=1$). The agent then enumerates the graphs $H$, and successively calls $\omeg$ with input $(2,H)$ until it gets a ``yes'' answer. When this occurs, the agent knows the quotient $Q$ of the graph in which it operates. The following claim shows that, in the case of a single agent, the class $\mav$ is stable with respect to the quotient
operation. For a node $u \in V(G)$, we denote by $\widehat{u}$ the equivalence class of $u$ in $\widehat{G}$.

\begin{claim}\label{claim:mav}
Let $\Pi\in\mav$ and let $(G,\{s\},i,w)$ be an initial configuration involving a single agent placed at node $s\in V(G)$. If $(G,\{s\},i,w)\in\Pi$ then, for every $H$ such that $\widehat{H}=\widehat{G}$,  and for every $t\in V(H)$ such that $\widehat{t}=\widehat{s}$, we have $(H,\{t\},i,w)\in \Pi$.
\end{claim}

To establish the claim, assume for the purpose of contradiction, that there exist two graphs $G$ and $H$, and two nodes $s\in V(G)$ and $t\in V(H)$ such that  $\widehat{H}=\widehat{G}$,  $\widehat{t}=\widehat{s}$, $(G,\{s\},i,w)\in\Pi$, and $(H,\{t\},i,w)\notin\Pi$ for some $i=\id(s)=\id(t)$ and some $w=\inp(s)=\inp(t)$. Let $\cP$ be a protocol verifying $\Pi$. By Lemma~\ref{lem:idexe}, given the same certificate $x$, the behavior of an agent starting at $s$ in $G$, and that of an agent starting at $t$ in $H$ are identical. Thus, for a certificate $x$ for which the agent decides ``yes'' for $(G,\{s\},i,w)$, it also decides ``yes'' for $(H,\{t\},i,w)$, a contradiction. This completes the proof of the claim. 

Claim~\ref{claim:mav} implies that the answer to the question whether $(G,\{s\},i,w)\in \Pi$ depends on $\widehat{G}$ rather than on $G$. So, to decide  whether $(G,S,\id,\inp)\in\Pi$, the agent has just to compute the quotient graph $\widehat{G}$ and  the equivalence class $\widehat{s}$ of the starting position $s$. By Claim~\ref{claim:view}, this can be achieved by computing the
truncated view from $s$ at depth $2(|V(\widehat{G})|-1)$. Once $\widehat{s}$  is identified, the agent can decide whether $(G,\{s\},i,w)\in\Pi$ by computing whether $(H,\{t\},i,w)\in\Pi$, where $H$ is an arbitrary graph such that $\widehat{H}=\widehat{G}$, and $t\in V(H)$ is an arbitrary node such that $\widehat{t}=\widehat{s}$. This completes the proof of Theorem~\ref{theo:main}. 
\end{proof}

%%%%%%%%%%%%%%%%%%%%%%%%%%%%%%%%%%%%%%%%%%%%%%%%%%%%%%%%%%%
\section{Beyond verifiability}
%%%%%%%%%%%%%%%%%%%%%%%%%%%%%%%%%%%%%%%%%%%%%%%%%%%%%%%%%%%

In this section we look beyond the class of mobile agents verifiable problems. What is the power of the information about the number of agents? We have seen in the proof of Theorem~\ref{theo:main} that it is very significant, if it is larger than~1. Indeed, if the agents know their number $k$, and this number is larger than 1, then they can compute both an isomorphic copy of the graph and the initial positions of all agents, up to automorphisms. Hence they can compute the entire initial configuration and consequently they can solve all decision problems in $\Delta$. By contrast, if the agent is only one (and even if it knows it), then the only problems it can solve are those from the class $\mad$. Hence, in order to study the power of additional information that can be provided to the agents, from now on we restrict attention to the case of a single agent. 
We  define the classes $\Delta_1$, $\mad_1$ and $\mav_1$ of decision problems, corresponding to $\Delta$, $\mad$ and $\mav$, respectively, but concerning a single agent. If $\cC$ is one of these classes, then $\cC_1$ is the class of problems $\Pi \in \cC$ whose instances are restricted to initial configurations $(G,S,\id,\inp)$ satisfying $|S|=1$. By Theorem~\ref{theo:main}, we get: 

\begin{corollary}
The decision problem $\quotient$ is $\mav_1$-complete.
\end{corollary}

For each decision problem $\Pi\in\Delta_1$, its complement $\overline{\Pi}\in\Delta_1$ is the decision problem such that, for every initial configuration $(G,\{s\},i,w)$, 
we have $(G,\{s\},i,w)\in \overline{\Pi}$ if and only if $(G,\{s\},i,w)\notin \Pi$. We then define $\mbox{co-}\mav_1$ in the standard way by: 
$$\Pi \in \mbox{co-}\mav_1 \iff \overline{\Pi}\in \mav_1~.$$
We have seen in the proof of Theorem~\ref{theo:main} that $\mav_1$ is stable with respect to the quotient operation, as a direct consequence of Lemma~\ref{lem:idexe}. By the same lemma, we get that $\mad_1$ and $\mbox{co-}\mav_1$ are also stable with respect to the quotient operation.

By definition, we have $\mad_1 \subseteq \mav_1 \cap \mbox{co-}\mav_1$. In fact, the two classes coincide. Indeed, consider $\Pi \in  \mav_1 \cap \mbox{co-}\mav_1$, and any initial configuration.  By enumerating all certificates $x\in\{0,1\}^*$, and, for each of them, using the verifying protocol $\cP_{\Pi}$ for the problem $\Pi$ and the verifying protocol $\cP_{\overline{\Pi}}$ for its complement, we get a  decision protocol for $\Pi$ because there is the first $x$ for which either $\cP_{\Pi}$ or $\cP_{\overline{\Pi}}$ will decide ``yes''. More precisely, for each tested certificate $x$, the agent applies $\cP_{\Pi}$ with certificate $x$. If the decision is ``no'', then the agent returns to its original position, and applies $\cP_{\overline{\Pi}}$ with the same certificate $x$. If the decision is again ``no'', then the agent returns to its original position, and picks the next certificate to be tested. It carries on that way until it gets a ``yes'' decision from $\cP_{\Pi}$ or $\cP_{\overline{\Pi}}$. If this ``yes'' is from $\cP_{\Pi}$ then it decides ``yes'', otherwise it decides ``no''. Hence, we get the following: 

\begin{theorem}
$\mav_1 \cap \mbox{\rm co-}\mav_1=\mad_1$.
\end{theorem}

Now, we move our concern to the universe beyond $\mav_1$ and $\mbox{\rm co-}\mav_1$, by considering decidability classes with oracles. Given a uniform decision problem $\Pi$,  the class $\mad ^{\Pi}$ is defined as the class of decision problems  $\Pi'$,
such that $\Pi' \preceq \Pi$, i.e., the class of problems that are mobile agents decidable, when an oracle deciding $\Pi$ is available to the agents.  
We are focussing attention on the classes $\mad_1^{\footnotesize \quotient}$, $\mad_1^{\footnotesize  \nbnode}$ and $\mad_1^{\footnotesize \map}$, where $\map=\{(G,S,\id,H): G= H\}$ is the problem to decide whether  the graph from the initial configuration is isomorphic to $H$,
for a given graph $H$ (as usual, an isomorphism has to preserve adjacencies and port numbers).
We want to compare these classes and situate them
with respect to  $\mad_1$ and  $\mav_1$. First notice that $\mad_1^{\footnotesize \quotient}$ is the class of problems for which the decision
may depend on the quotient graph $\widehat{G}$ of the graph $G$ from the initial configuration, 
 but not on the graph $G$ itself.  Indeed, having the oracle $\quotient$ is equivalent, for a single agent, to having a copy of $\widehat{G}$:
the agent can query this oracle with all arbitrarily enumerated graphs, until the answer is ``no''. Similarly, the oracle $\nbnode$ is equivalent to having the size of the 
underlying graph, and the oracle $\map$ is equivalent to having an isomorphic copy of the underlying graph.
We show that providing a single agent operating in the graph  $G$ with a copy of the quotient graph
$\widehat{G}$, permits it to solve all mobile agents verifiable and co-verifiable problems, as well as some problems outside of these classes. This is implied by the following theorem.

\begin{theorem}\label{incl1}
$\mad_1 \subset (\mav_1 \cup \mbox{\rm co-}\mav_1) \subset \mad_1^{\footnotesize \quotient}$ (all inclusions are strict).
\end{theorem}

\begin{proof}
The inclusion $\mad_1 \subseteq \mav_1 \cup \mbox{\rm co-}\mav_1$ is by definition. It is strict, as witnessed, e.g., by the  problem $\path$. $\mav_1 \cup \mbox{\rm co-}\mav_1 \subseteq \mad_1^{\footnotesize \quotient}$ by Lemma~\ref{lem:idexe}. We show that the inclusion is strict. Let $\cycle=\{(G,S,\id,\epsilon):G\;\mbox{is a cycle}\}$ and $\sun=\{(G,S,\id,\epsilon): G \;\mbox{is a sun}\}$ where Figure~\ref{fig:proof}(a) displays a sun. (In both cases, we assume that the edges are consistently labeled~1 clockwise, and~2 counterclockwise). We have $\cycle\times\overline{\sun} \notin \mav_1 \cup \mbox{\rm co-}\mav_1$. Indeed,  $\cycle\notin\mav_1$ and $\sun\notin\mav_1$. On the other hand, both problems $\cycle$ and $\overline{\sun}$ are in $ \mad_1^{\footnotesize \quotient}$. Indeed, let $O$ be the graph consisting of a unique node with a loop  labeled~1 and~2 at its two extremities, and let $P$ be the graph consisting of two nodes $u$ and $v$ connected by an edge, with a loop at node $u$ labeled~1 and~2 at its two extremities. We have $\widehat{G}=O$ if and only if $G$ is a (consistently labeled) cycle, and $\widehat{G}=P$ if and only if $G$ is a (consistently labeled) sun. 
\end{proof}

We conclude with the following result that compares classes  $\mad_1^{\footnotesize \quotient}$, $\mad_1^{\footnotesize\nbnode}$ and $\mad_1^{\footnotesize\map}$.
It shows that, in the case of a single agent, providing the size of the graph is stronger than providing the quotient graph, and providing an isomorphic copy
of the graph is even stronger and permits the agent to reconstruct its initial position (up to authomorphism), i.e., it is as strong as providing the entire initial configuration.

\begin{theorem}\label{incl2}
$\mad_1^{\footnotesize \quotient}\subset \mad_1^{\footnotesize\nbnode} \subset \mad_1^{\footnotesize\map} = \Delta_1$ (all inclusions are strict). 
\end{theorem}

\begin{proof}
The inclusion chain follows from the facts that knowing the size of a graph allows an agent to compute its quotient graph (by Theorem~\ref{trunc}), and  knowing the map of a graph allows an agent to compute its size. The inclusions are strict. Indeed, $\nbnode\notin \mad_1^{\footnotesize \quotient}$, and $\map\notin\mad_1^{\footnotesize\nbnode}$. 
The former holds because there exist graphs of different sizes with the same quotient (e.g., cycles), and
the latter holds because there exist non-isomorphic graphs with the same size and the same quotient (see, e.g., the two graphs displayed in Figure~\ref{fig:proof}~(c) and~(d), cf. \cite{YK3}). Finally, the fact that $\mad_1^{\footnotesize\map}$ is actually equal to the set $\Delta_1$ of all decidable problems involving configurations with a single agent follows from the fact that if the map $G$ is given to the agent then it can compute its starting position $s\in V(G)$ (up to automorphism) on the map, and thus it can compute the whole initial configuration $(G,\{s\},i,w)$. 
\end{proof}

\begin{figure}[tb]
\begin{center}
\includegraphics[width=9cm]{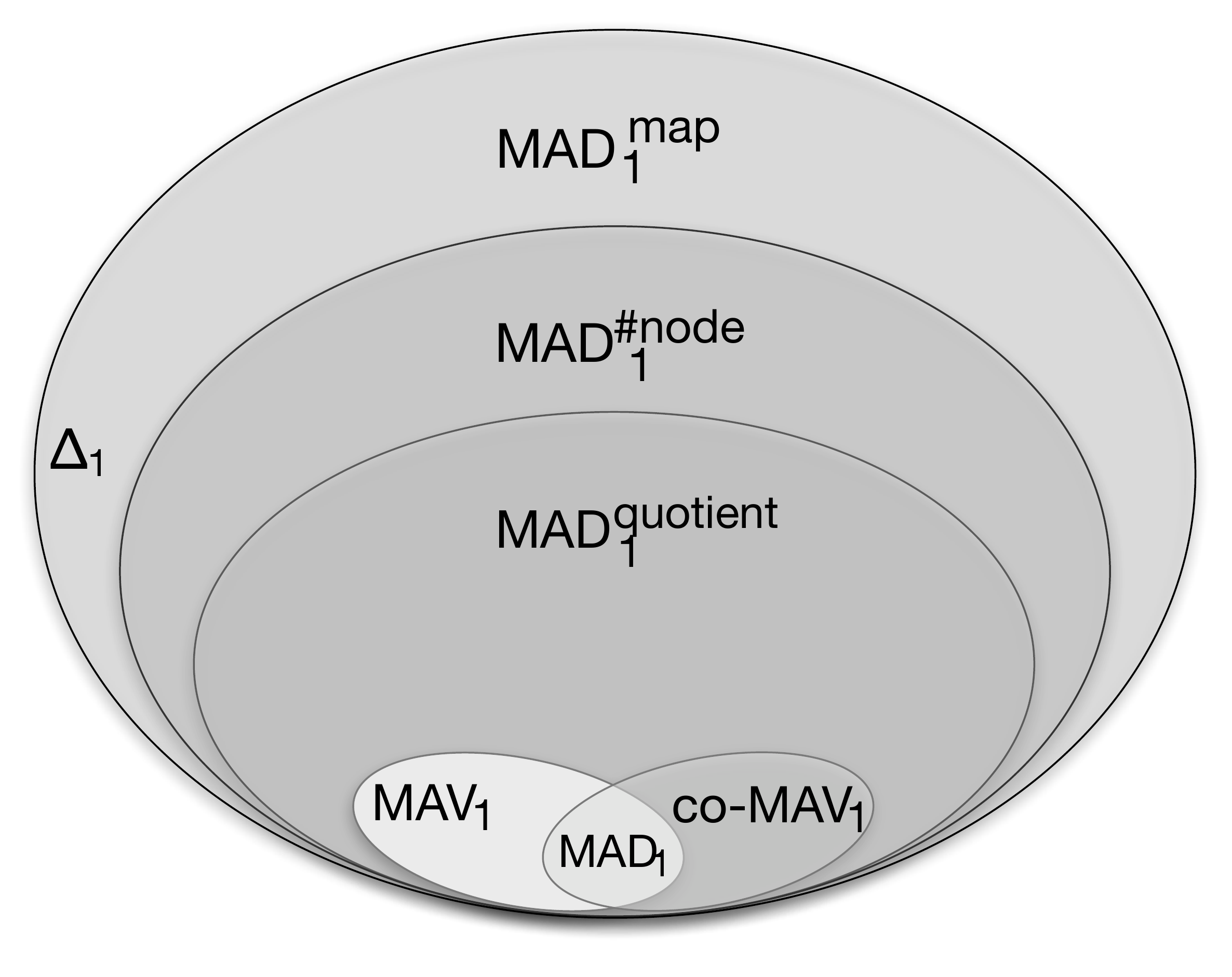}
\caption{Structure of the class $\Delta_1$ of decision problems for a single agent}
\label{fig:summary}
\end{center}
\end{figure}

Figure~\ref{fig:summary} summarizes all the results in this section. 

%%%%%%%%%%%%%%%%%%%%%%%%%%%%%%%%%%%%%%%%%%%%%%%%%%%%%%%%%%%
\section{Conclusion and open problems}
%%%%%%%%%%%%%%%%%%%%%%%%%%%%%%%%%%%%%%%%%%%%%%%%%%%%%%%%%%%

We provided a classification of decision problems that can be solved by mobile agents operating in graphs, possibly with the help of some additional information.
It turns out that knowing the size of the team of agents is very powerful but only if this size is larger than 1. In this case, knowing the size of the team permits the agents
to gather and then reconstruct the underlying graph and the entire initial configuration, i.e., they can solve all decision problems. This points to the crucial role of gathering
(or rendezvous) in mobile agents computing. It is also an additional justification of the importance of this task which gained growing attention of the algorithmic community in the recent years. The following question remains open: what is the minimal information about the size of the team of agents (if there are many of them) that permits
the agents to reconstruct the initial
configuration? 

For a single agent, our three classes of relative mobile agents decidability, $\mad_1^{\footnotesize \quotient}$, $\mad_1^{\footnotesize\nbnode}$ and $\mad_1^{\footnotesize\map}$, seem to be fairly natural,  as they correspond to types of information
that are often given to agents accomplishing some task in a communication network. The size of the graph or its topology are examples of such information. 
The quotient graph is rarely given to the agent but its importance comes from the fact that this is the ultimate information about the underlying graph that the agent
can acquire when operating in it.
It would be interesting to have a combinatorial characterization of the classes $\mad_1^{\footnotesize \quotient}$, $\mad_1^{\footnotesize\nbnode}$ , as well as
of the class $\mad_1$ itself, that would permit to decide if a given problem is in a particular class.
It seems particularly intriguing if classes  $\mad_1$ and $\mad_1^{\footnotesize\nbnode}$ are decidable.
In other words, is it possible to decide if a given decision problem $\Pi\in\Delta_1$ can be solved by a mobile agent operating in this graph (not having any
a priori information), or, alternatively, by an agent just knowing the size of the graph? 
A more precise formulation of this open problem is the following. Does there exist a Turing machine which, given as input a description of a Turing machine deciding if an initial 
configuration is in $\Pi$, decides whether $\Pi$ is in $\mad_1$ (respectively in $\mad_1^{\footnotesize\nbnode}$).

\newpage

%%%%%%%%%%%%%%%%%%%%%%%%%%%%%%%%%%%%%%%%%%%%%%%%%%%%%%%%%%%
\bibliographystyle{plain}

%%%%%%%%%%%%%%%%%%%%%%%%%%%%%%%%%%%%%%%%%%%%%%%%%%%%%%%%%%% 

\end{document}